\documentclass[aps,showpacs,twocolumn,longbibliography,nofootinbib,secnumarabic,superscriptaddress]{revtex4-1}
\usepackage{graphicx}
\usepackage{color}
\usepackage{amsmath}
\usepackage{epstopdf}

\begin{document}

\title
{Single-mode nonclassicality criteria via Holstein-Primakoff transformation}
\author{Mehmet Emre Tasgin}
\affiliation{Institute of Nuclear Sciences, Hacettepe University, 06800, Ankara, Turkey}
\email{metasgin@hacettepe.edu.tr}
\email{metasgin@gmail.com}

\date{\today}

\begin{abstract}
We study different representations of a many-particle system. We reveal the stunning beauty of the connections among the nonclassicalities of a many-particle, a single-mode and a two-mode systems. First, we demonstrate that (separable) atomic coherent states~(ACSs), of a many-particle system, mimic the coherent states of single-mode light. This enables the utilization of a many-particle entanglement~(MPE) criterion as a single-mode nonclassicality~(SMNc) condition via Holstein-Primakoff transformation. As an example, we show that spin-squeezing criterion, for MPE, can be transformed into the quadrature-squeezing, a nonclassicality condition for a single-mode system. Second, we demonstrate a similar connection between the two-mode separable states and separable ACSs of a many-particle system. The second connection makes us be able to obtain a single-mode nonclassicality condition from a two-mode entanglement criterion, using the many-particle system as a bridge. As an example, we obtain Mandel's Q-parameter from the two-mode entanglement criterion of Hillery{\&}Zubairy.
\end{abstract}



\maketitle

\section{Introduction}

Nonclassical quantum states are the basis for quantum technologies which are believed to reshape the new century~\cite{quantumflagship}. Technologies like quantum teleportation~\cite{BraunsteinNaturePhot2015}, measurements below the standard quantum limit~(SQL)~\cite{MeasurementSQLSciene2004,LIGO2013}, quantum radars~\cite{QuantumRadarSciRep2017}, quantum computation~\cite{Nielsen:10} etc., all, require nonclassical states. Hence, detection and quantification of the nonlcassicality of such states possess fundamental importance in the field.

Although nonclassicality may appear in different forms, they actually can be converted into each other. Single-mode nonclassicality~(SMNc) of an incident light mode, e.g. squeezing in quadrature or number fluctuations, can be converted into two-mode entanglement~(TME) at the output of a beam splitter~(BS). Moreover, a conservation-like relation holds between single-mode nonclassicality and two-mode entanglement in a beam splitter~\cite{ge2015conservation,arkhipov2016nonclassicality,arkhipov2016interplay,vcernoch2018experimental}, a passive device which is unable to generate extra nonclassicality~\cite{Kim:02}. That is, not all of the SMNc of the incident (input) mode can be converted into TME in the beam splitter, but some SMNc remains in both output modes. A single-mode light can also transfer its nonclassicality~(in the form of SMNc), e.g. squeezing, into an ensemble of atoms~\cite{PolzikPRL1999spinsqz,TothPRA2018SpinSqz,vidal2006concurrence}. Quadrature-squeezing of the incident light transforms into spin-squeezing~\cite{kitagawa1993squeezed} in the ensemble. Spin-squeezing creates a collective (many-particle, MP) entanglement~(MPE) between the constituent atoms~\cite{sorensen2001Nature,tasgin2017many}. Similarly, interaction of an entangled two mode light with an ensemble creates MPE in the ensemble~\cite{regula2018converting,Tasgin&MeystrePRA2011}. Hence, nonclassicality~(Nc) is like ``energy": it can be transformed into different forms. 

Not only different forms of nonclassicalities can be converted into each other, but also criteria witnessing different forms of nonclassicalities can be transformed into each other. For instance, SMNc criteria can be obtained from TME criteria using a beam splitter transformation, see Refs.~\cite{hillery2006PRA,tasgin2015measure,tasgin2015HP} and Eqs.~(\ref{a1a2BS1}-\ref{a1a2BS2}) in the present paper. A single-mode nonclassical state, put into a beam splitter, generates entanglement between the output modes~\cite{Kim:02,Asboth:05}. This makes a TME criterion work as a SMNc criterion~\cite{hillery2006PRA,tasgin2015measure,tasgin2015HP} via entanglement potential~\cite{Asboth:05}. 

In Ref.~\cite{tasgin2017many}, we further demonstrate (i) how one can obtain a SMNc criterion from a MPE criterion and (ii) how one can anticipate a new MPE criterion from the form of a SMNc criterion, e.g. a new MPE criterion from Mandel's Q parameter. Such a connection appears because an atomic coherent state~(ACS)~\footnote{Also known as spin coherent states.} of an ensemble, a separable state~\cite{MandelWolfbook}, becomes the coherent state of a single-mode light, a classical state, for $N\to\infty$~\cite{klauder1985applications,radcliffe1971JPhysA}. Ref.~\cite{gunay2019entanglement}, in a similar manner, shows that some forms for ensemble-ensemble entanglement criteria can be anticipated from the structure of TME criteria.

Quantum state of an ensemble of identical (indistinguishable) particles has to be in a symmetric form with respect to the exchange of particles. In this case, the dynamics of the system can be equivalently described via operators $\hat{c}_g$ and $\hat{c}_e$ which annihilates/creates a particle in the ground and excited states, respectively~\cite{ScullyZubairyBook}. In such a system, a connection between the entanglement of these two modes, $\hat{c}_g$--$\hat{c}_e$, and the entanglement among the atoms of the ensemble (a many-particle entanglement) has been intrigued in Refs.~\cite{dalton2014NJP,dalton2017quantum}. More specifically, Ref.~\cite{dalton2014NJP} shows that the spin-squeezing~(MPE) criterion~\cite{sorensen2001Nature} cannot be satisfied unless the two-modes describing this $N$-particle two-level system are entangled.

 As a counter-example, for the general case, however, it is straightforward to show that a separable two mode state, e.g., $|n_g\rangle_g |n_e\rangle_e$ demonstrates a strong many-particle entanglement. Here, $n_g$ and $n_e$ denote the occupation of the modes associated with the ground and excited states.  $|n_g\rangle_g |n_e\rangle_e$ is actually the mode representation of a Dicke state~\cite{MandelWolfbook}, where $n_e$  number of excitation is equally and symmetrically distributed among $N=n_g+n_e$ number of atoms. $|n_{g,e}\rangle_{g,e}$ are the Fock states. Besides the two-mode representation, a Dicke state can as well be represented by spin states, e.g. as $|S,m\rangle$ with $S=N/2=(n_g+n_e)/2$ and $m=(-S+n_e)$~\cite{MandelWolfbook}. The counter-example demonstrates that many-particle entanglement, in an ensemble of $N$ identical two-level particles, is possible to emerge via presence of a single-mode nonclassicality in one of the two modes, besides the two-mode entanglement.

In this paper, we investigate the connection between single-mode nonclassicality, two-mode entanglement and many-particle entanglement criteria in the limit $N\to\infty$. 
(a) We show that in the $N\to\infty$ limit, entanglement of the modes $\hat{c}_g$--$\hat{c}_e$ is \textit{sufficient} for the presence of the (collective) many-particle entanglement of the particles in the ensemble. In demonstrating this connection, it gets required to show the absence of single-mode nonclassicality in both of the modes, $\hat{c}_g$ and $\hat{c}_e$, while taking the limit $N\to\infty$. (b) Next, we use this, TME$\to$MPE, connection to be able obtain a single-mode nonclassicality criterion from a two-mode entanglement criterion. This  provides a new method, alternative to the beam splitter approach~\cite{hillery2006PRA,tasgin2015measure,tasgin2015HP}, for obtaining single-mode nonclassicality conditions from two-mode entanglement criteria. Obtaining an analytical form does not always become possible via beam splitter approach. (c) What is more important is; our study brings a clarification to the connections between nonclassicalities associated with many-particle, single-mode and two-mode systems. (d) In Ref.~\cite{tasgin2017many}, we have already demonstrated that a many-particle entanglement criterion becomes a single-mode nonclassicality condition for $N\to\infty$. Here, for completeness, in Sec.~\ref{sec:spin-squeezing}, we additionally demonstrate that spin-squeezing criterion for many-particle entanglement~\cite{sorensen2001Nature} becomes the quadrature-squeezing condition, a SMNc criterion.

Our approach, despite presenting an intriguing TME-SMNc connection and can lead to a better anticipation of MPE criteria, has a drawback. It can transform the TME criteria into SMNc condition only when, at least in the present form, the TME criterion involves particle number conserving terms, e.g. like $\hat{c}_e^\dagger\hat{c}_g$ but not like $\hat{c}_e\hat{c}_g$.

Two-mode entanglement criteria are more in number compared to the MPE and SMNc criteria. Hence, an approach similar to Ref.~\cite{tasgin2017many} can be used to anticipate a form for a MPE criterion from the form of a TME criterion. For a better visualization: Ref.~\cite{tasgin2017many} anticipates one of the possible forms for a MPE criterion from the SMNc condition Mandel's $Q$-parameter, by examining~\cite{sorensen2001Nature} the uncertainty of the operator $\hat{\cal R}=\hat{S}_+\hat{S}_-$ which becomes $\hat{n}=N\hat{b}^\dagger\hat{b}$ in the $N\to\infty$ limit via Holstein-Primakoff transformation. A given TME criterion can also be used for (a detailed) anticipation of a form for a MPE criterion.


The paper is organized as follows. Derivations of the connections MPE$\to$SMNc and TME$\to$MPE are intimately related to each other. So, for the coherence, we demonstrate the connections among the three systems in Sec.~\ref{sec:connection} and present our examples in Sec.~\ref{sec:examples}. In Sec.~\ref{sec:ACS} we overview the atomic coherent states~(ACSs), separable many-particle states, and the Dicke states, highly entangled many-particle states. In Sec.~\ref{sec:MPEtoSMNc}, we present the spin representation for Dicke states and ACSs, and show that they mimic the Fock and coherent states of light~\cite{klauder1985applications,radcliffe1971JPhysA}, respectively, as number of the particles becomes very large $N\to\infty$. In Sec.~\ref{sec:spin-squeezing}, we show that spin-squeezing (a MPE) criterion becomes the quadrature-squeezing (a SMNc) condition. In Sec.~\ref{sec:TMrepresentation}, we introduce the two mode representation of a many-particle system. In Sec.~\ref{sec:TMEtoMPE} we show that presence of entanglement between the two modes, $\hat{c}_g$ and $\hat{c}_e$, of the ensemble implies the presence of many-particle entanglement in the $N\to\infty$ limit. Here, in difference to Ref.~\cite{tasgin2017many}, we face to show that an ACS does not possess nonclassicality in both modes. Presence of a many-particle entanglement in the ensemble of this identical particles, however, does not imply the presence of entanglement even in the $N\to\infty$ limit. In Sec.~\ref{sec:QfromH&Z}, we demonstrate how one can obtain a SMNc condition from a TME criterion using the path TME$\to$MPE$\to$SMNc. As an example, we obtain the  Mandel's $Q$-parameter ($Q<1$) condition from Hillery-Zubairy criterion~\cite{Hillery&ZubairyPRL2006}, where both work good for number-squeezed like states~\cite{NhaPRA2006Fock_states}. In Sec.~\ref{sec:Q_withBS}, for completeness, we also provide the alternative definition of Mandel's $Q$-parameter (a SMNc condition) from the Hillery{\&}Zubairy TME criterion, this time, using the beam splitter approach. Sec.~\ref{sec:conclusions} contains  our summary and discussions.

\section{Connections between many-particle, single-mode and two-mode systems} \label{sec:connection}

\subsection{Atomic and single-mode coherent states} \label{sec:ACS}

Quantum state of an ensemble of two-level particles can be represented in several ways. One can, simply, write the ensemble state in terms of single particle states, $|g\rangle_i$ or $|e\rangle_i$, where $i$th particle is in the ground and excited state. Alternatively, the ensemble state can be represented by angular momentum addition theorem for $N$ spin-1/2 particles~\cite{MandelWolfbook}, called as Dicke states, see figure 1 in Ref.~\cite{dicke1954SR}. Below, we refer to both representations for gaining understanding on the nonclassical properties of a many-particle system. 

Ground state of such an ensemble, all particles are in the ground state, can be written as
\begin{equation}
|\psi_N^{\rm \scriptscriptstyle (ground)}\rangle=|S,-S\rangle=|g\rangle_1 \otimes |g\rangle_2 \ldots \otimes |g\rangle_N.
\label{ACSgs}
\end{equation} 
Here, total-spin is $S=N/2$. $|S,m\rangle$, with $m=-S\ldots S$, are called as Dicke states and are highly (collectively) many-particle entangled for $m\neq\pm S$. According to the addition of angular momentum, $S$ can assign values from $0\ldots N/2$. Here, however, we deal only with (exchange) symmetric set of Dicke states where $S=N/2$. Dicke states are usually employed in  describing collective phenomena in ensembles such as superradiance \cite{emary2003chaos,lambert2005entanglement}.  When the particles are identical, either bosons \cite{Ketterle1999Science_superradiant} or fermions \cite{fermionicSR2011}, only the symmetric set (maximum cooperation $r=N/2$) of Dicke (internal) states can be occupied~\cite{symmetricDicke,PS_BECSRvsSR}. 

In order to represent the occupation of the excited state, one can equivalently represent a (symmetric)  Dicke state as $|S,m\rangle \equiv |N,n_e\rangle$ where $n_e$ is the occupation of the excited state. Here $n_e=S+m$ and for $m=-S$ all particles are in the ground state. The Dicke~\footnote{In the remaining part of the ext, we refer to symmetric Dicke states.} states become the well-known Fock states of single-mode light $|N, n_e\rangle \to |n_e\rangle$ in the limit $N\to\infty$~\cite{tasgin2017many,klauder1985applications,radcliffe1971JPhysA}. 

An atomic coherent state~(ACS) $|S,z\rangle$ can be generated from the ground state $|S,-S\rangle$ by applying the collective atomic displacement operator
\begin{equation}
\hat{D}_a(\xi)=e^{\xi \hat{S}_+ -\xi^*\hat{S}_-} \quad \text{as}
\label{Da}
\end{equation}
\begin{eqnarray}
&&|S,z\rangle=\hat{D}_a |S,-S\rangle  \nonumber \\
&&=\frac{1}{(1+|z|^2)^{N/2}} \sum_{m=-S}^S \dbinom{N}{S+m}^{1/2} z^{S+m} |S,m\rangle,
\label{ACSgenerate}
\end{eqnarray}
where $\hat{S}_\pm$ are ladder operators with 
\begin{equation}
\hat{S}_\pm |S,m\rangle =\left[(S\mp m)(S\pm 1)\right]^{1/2} |S,m\pm 1\rangle
\label{ladder}
\end{equation}
and $z=\tan|\xi|e^{i{\rm arg}\{\xi\}}$~\cite{MandelWolfbook}.

An atomic coherent state has a special property. It is a separable state, that is, it can be written as a product of single atom states 
\begin{equation}
|S,z\rangle=\text{\cal C}\big(|g\rangle_1+z|e\rangle_1\big) \otimes  \ldots \otimes \big(|g\rangle_N+z|e\rangle_N\big),
\label{ACS}
\end{equation}
similar to Eq.~(\ref{ACSgs}), where $\text{\cal C}$ stands for the normalization constant $\text{\cal C}=1/(1+|z|^2)^{1/2}$. Hence, an atomic coherent state is a subset of mixed separable states 
\begin{equation}
\hat{\rho}=\sum_{k}P_k\;\hat{\rho}_k^{(1)}\otimes\hat{\rho}_k^{(2)}\otimes\ldots\otimes\hat{\rho}_k^{(N)} ,
\label{SorensenEnt}
\end{equation}
studied, e.g., by S{\o}rensen {\it et al.}~\cite{sorensen2001Nature,sorensen2001entanglement}.

In the large number of particles limit $N\to\infty$, Dicke states transform to Fock states of a single-mode light, i.e. $|S,m\rangle\equiv |N,n_e\rangle\to |n_e\rangle$. When the summation in Eq.~(\ref{ACSgenerate}) is carried over an infinite number of particles, the coefficients of the ACS, $|S,z\rangle$, converge to $\propto \alpha^n/\sqrt{n!}$~\cite{klauder1985applications,radcliffe1971JPhysA,tasgin2017many}, to the coefficients of the well-known coherent states of light. 

Therefore, an atomic coherent state transforms to coherent state of single-mode light at the infinite-$N$ limit, i.e.
\begin{equation}
|S=N/2,z\rangle \to |\alpha\rangle.
\end{equation}
Below, using this phenomenon we obtain a connection between the strengths of many-particle entanglement in an ensemble and nonclassicality of a single-mode light. 

\subsection{Many-particle entanglement and \\ single-mode nonclassicality} \label{sec:MPEtoSMNc}

An $N$-particle state of indistinguishable particles can be written as a superposition of ACSs, in the angular momentum representation, 
\begin{equation}
|\psi_N\rangle = \sum_{i=1}^r \kappa_i \; |N/2,z_i\rangle \; ,
\end{equation}
or equivalently in terms of separable (product) single particle states~\cite{sorensen2001Nature}
\begin{equation}
|\psi_N\rangle = \sum_{i=1}^r \kappa_i \; \left( |g\rangle_1+z_i|e\rangle_1\right) \otimes \ldots \otimes \left( |g\rangle_N+z_i|e\rangle_N\right) .
\end{equation}
Here, both rank $r$ and the coefficients $\kappa_i$ determine the nonclassical features of both the many-particle system and the single-mode system~\cite{tasgin2017many} which becomes
\begin{equation}
|\psi\rangle=\sum_{i=1}^r  \kappa_i \; |\alpha_i\rangle.
\label{rcoh}
\end{equation}
 
Therefore, strength of the many-particle entanglement in $|\psi_N\rangle$ determines also the strength of the nonclassicality of the single-mode field it maps, as $N\to\infty$. This observation can be used for deriving a SMNc condition from a given MPE criterion~\cite{tasgin2017many}. 

Nonclassicality (all MPE, TME and SMNc) criteria, however, adapt the expectation values of observables: the operators. So, for our connection 
\begin{equation}
\sum_{i=1}^r \kappa_i \; |N/2,z_i\rangle \xrightarrow{N\to \infty} \sum_{i=1}^r  \kappa_i \; |\alpha_i\rangle
\end{equation}
to be useful, we need to establish a relation also between the operators belonging to the many-particle system, e.g. $\hat{S}_{\pm,z}$, and the ones belonging to the single-mode system $\hat{a}$ and $\hat{a}^\dagger$ (or $\hat{b}$ and $\hat{b}^\dagger$). The operators for the two systems can also be mapped via Holstein-Primakoff transformation.

\subsection{Holstein-Primakoff transformation} \label{sec:HPtransform}

The operators of a symmetric many-particle system and a single-mode system can be related via Holstein-Primakoff transformation~\cite{emary2003chaos,Holstein&Primakoff}  
\begin{eqnarray}
&&\hat{S}_+=\hat{b}^{\dagger} \sqrt{N-\hat{b}^{\dagger}\hat{b}} \quad , \quad \hat{S}_-=\sqrt{N-\hat{b}^{\dagger}\hat{b}} \; \hat{b} \nonumber 
\\
&&{\rm and} \quad \hat{S}_z=\hat{b}^{\dagger}\hat{b}-N/2
\label{HPtransform}
\end{eqnarray}
without referring to states explicitly. Here, for finite $N$, we distinguish the operator $\hat{b}$ from the annihilation operator $\hat{a}$ of a single-mode light. When $N$ is sufficiently large, the ladder operators transform to single-mode annihilation/creation operators
\begin{equation}
\hat{S}_+ \rightarrow \sqrt{N}\hat{b}^{\dagger} \quad , \quad \hat{S}_- \rightarrow \sqrt{N}\hat{b} 
\quad {\rm and} \quad \hat{S}_z \rightarrow -N/2 +  \hat{b}^{\dagger} \hat{b} \; ,
\end{equation}
where, now, $\hat{b}$ stands for $\hat{a}$ of a single-mode state, i.e. $\hat{a}\equiv\hat{b}$. (In the following subsection, we also show that in the two-mode representation it is $\hat{b}\equiv \hat{c}_e$.) In Sec.~\ref{sec:spin-squeezing},using the maps introduced in Secs.~\ref{sec:MPEtoSMNc} and \ref{sec:HPtransform}, we demonstrate that quadrature-squeezing~(a SMNc) condition can be obtained from spin-squeezing~(a MPE) criterion~\cite{sorensen2001Nature}.

It is worth also noting that collective atomic displacement operator $\hat{D}_a(z)$, in Eq.~(\ref{Da}), transforms to single-mode displacement operator
\begin{equation}
\hat{D}(\beta)=e^{\beta\hat{b}^\dagger-\beta^*\hat{b}}
\end{equation}
 for $N\to\infty$. There is a striking common feature between the ensemble operator $\hat{D}_a(\xi)$ (subscript`$a$' stands for atoms) and the displacement operator for single-mode light $\hat{D}(\beta)$, with $\beta=\sqrt{N}\xi$. In generating the separable ACSs, in Eqs.~(\ref{ACSgenerate}) and (\ref{ACS}), from the separable ground state, in Eq.~(\ref{ACSgs}), $\hat{D}_a(\xi)$ does not change the nonclassicality features of the state it operates on. Both initial and final states are separable many-particle states. Similarly, as well-known from quantum optics courses~\cite{ScullyZubairyBook}, the displacement operator $\hat{D}(\beta)$ does not change the noise features of the single-mode system~\cite{simon1994quantum}. In the following subsection this relation is enriched with an extension to the two-mode system.

\subsection{Two-mode representation of a many-particle system} \label{sec:TMrepresentation}

Another (a third) representation for Dicke states, alternative to single-particle states $|e\rangle_i$ or total-spin states $|S,m\rangle$, is the two-mode representation $|n_g,n_e\rangle$ with the particle number super-selection~\cite{dalton2014NJP} constraint $n_g+n_e=N$. In the Holstein-Prikamoff transformation, in Eq.~(\ref{HPtransform}), actually, $\hat{b}\equiv\hat{c}_e$ and $\sqrt{N-\hat{b}^\dagger\hat{b}}\equiv \hat{c}_g$. In the infinite $N$ limit of the Holstein-Primakoff transformation, we actually replace $\hat{c}_g\to\sqrt{N}$, as in parametric approximation, where the reduction in the number of, e.g., laser photons is negligibly small. Here, this corresponds to: number of particles is infinite so a finite $n_e$ alters the $n_g$ from $N$ only negligibly. 

In the two-mode representation, the operator $\hat{D}_a(\xi)$, in Eq.~(\ref{Da}), which generates the atomic coherent states in Eq.~(\ref{ACS}), becomes
\begin{equation}
\hat{D}_{\rm \scriptscriptstyle TM}(\xi)= e^{\xi\hat{c}_e^\dagger\hat{c}_g - \xi^*\hat{c}_e^\dagger\hat{c}_g},
\end{equation}
which is simply a beam splitter operator~\cite{ge2015conservation}. Here subscript TM stands for two-mode representation. It is known that a beam splitter cannot generate entanglement between the two output modes unless the two input modes possess nonclassicality, see Refs.~\cite{Kim:02} and \cite{ge2015conservation,tasgin2019anatomy,arkhipov2016nonclassicality,arkhipov2016interplay,vcernoch2018experimental}. It merely transforms the SMNc into TME, or vice versa. 

Therefore, all three representations of the $\hat{D}_a(\xi)$ operator do not create nonclassicality in the single-mode, two-mode and many-particle states they act on. This consolidates the beauty of the connections among the three representations.

\subsection{Two-mode entanglement and \\ Many-particle inseparability} \label{sec:TMEtoMPE}

In this subsection, we show that a two-mode entanglement criterion can be transformed into a single-mode nonclassicality condition by using the many-particle system as a bridge. First, we transform a TME criterion into a many-particle entanglement criterion. Then, using the method introduced in Secs.~\ref{sec:MPEtoSMNc} and \ref{sec:HPtransform} and in Ref.~\cite{tasgin2017many}, we obtain a SMNc condition via Holstein-Primakoff transformation.

We follow the the same strategy we do in Sec.~\ref{sec:MPEtoSMNc}. We show that a superposition of atomic coherent states can be written as a superposition of separable two-mode states. Doing this, however, one needs to be careful. Because an intra-mode nonclassicality, existing in one of the two modes, may also be responsible for the many-particle entanglement in the mapping.

We aim to show
\begin{equation}
|\psi\rangle_N = \sum_r |\xi_{\rm \scriptscriptstyle ACS}^{(r)}\rangle \xrightarrow{N\to\infty} |\psi\rangle=\sum_r |\ldots\rangle_g\otimes|\alpha^{(r)}\rangle_e
\label{MPEtoTME}
\end{equation}
in the infinite particle number limit. All ACSs are generated from the symmetric ground state $|\psi_N^{(\rm \scriptscriptstyle gound)}\rangle=|S,-S\rangle=|n_g=N\rangle|n_e=0\rangle$, as $|\xi_{\rm \scriptscriptstyle ACS}^{(r)}\rangle=e^{\hat{D}_a}|N\rangle|0\rangle$, see Eq.~(\ref{ACSgenerate}). Hence, before all, first we need to consider the nonclassicality of $|\psi_N^{(\rm gound)}\rangle$. (i) The state $|n_g=N\rangle|n_e=0\rangle$ is obviously separable. That is, a nonclassicaity associated with TME does not exist in this state. (ii) The second mode, $|n_e=0\rangle_e$, also does not possess any single-mode nonclassicality. (iii) However, we need to be sure (show) whether $|n_g=N\rangle_g$, a Fock state for finite $N$, does not possess SMNc as $N\to \infty$. That is, we also need to show $|\xi_{\rm \scriptscriptstyle ACS}^{(r)}\rangle \xrightarrow{N\to\infty}  |\ldots\rangle_g\otimes|\alpha^{(r)}\rangle_e$ does not possess any nonclassicality. Actually, this [i.e. (iii)] is the main step in demonstrating $\xi_{\rm \scriptscriptstyle ACS} \xrightarrow{N\to\infty} |\ldots\rangle_g\otimes |\alpha\rangle_e$. 

Such a relation can be demonstrated safely as follows. $\hat{D}_a=e^{\xi \hat{S}_+ -\xi^*\hat{S}_-}\equiv e^{\xi\hat{c}_e^\dagger\hat{c}_g - \xi^*\hat{c}_e^\dagger\hat{c}_g}$, generating the ACSs, is a beam splitter transformations, even for finite $N$. If $|n_g=N\rangle|n_e=0\rangle$, from which all ACSs are generated via $\hat{D}_a$, or the atomic coherent states themselves $|\xi_{\rm \scriptscriptstyle ACS}\rangle_{N\to \infty}$ had demonstrated any kind of nonclassicality, i.e. TME or SMNc; then at some value of $\xi$ there would exist SMNc in $|\alpha\rangle_e$. Because, $\hat{D}_a$ is a beam splitter operator and a beam splitter operator transforms TME and SMNc in a two-mode system~\cite{ge2015conservation}. However, we know that in the $N\to\infty$ limit the second mode does not possess any SMNc~\cite{klauder1985applications,radcliffe1971JPhysA}, \cite{tasgin2017many}. Hence, none of the two modes ($\hat{c}_{g,e}$), representing $|\xi_{\rm \scriptscriptstyle ACS}\rangle_{N\to \infty}$, contains any nonclassicality. This further proves that $\xi_{\rm \scriptscriptstyle ACS} \xrightarrow{N\to\infty} |\ldots\rangle_g\otimes |\alpha\rangle_e$ possesses no nonclassicality at all.

Therefore, if there is TME between $\hat{c}_g$--$\hat{c}_e$, for $N\to\infty$, then there is MPE in $|\psi_N\rangle$. In other words, when the two-mode state in Eq.~(\ref{MPEtoTME}) cannot be written as a single term, i.e. $r \neq 1$, then the many-particle state is also a superposition of ACSs. Hence, the many-particle state is entangled for $N\to\infty$.

Using this relation, we can also derive a SMNc condition from a TME criterion tracking the following path. (i) We can transform $\hat{c}_{g,e}$ in the TME criterion, e.g., as $\hat{c}_e^\dagger\hat{c}_g=\hat{S}_+$ which checks whether there is MPE in the many-particle system. (ii) Then, we can make a Holstein-Primakoff transformation as $N\to\infty$: $\hat{S}_+\to\sqrt{N}\:\hat{b}^\dagger$. (iii) Finally, we obtain a SMNc from the TME. This is an alternative to beam splitter transformation which can be used to derive SMNc criteria from TME criteria~\cite{tasgin2015measure}.

We note that a MPE criterion cannot be used for detecting the TME. Because, the nonclassiclality, MPE demonstrates, could be also due to single-mode nonclassicality of the one of the two modes, as well as it could originate from the entanglement between the two modes.

In Sec.~\ref{sec:QfromH&Z}, as a demonstration of the method, we obtain Mandel's Q-parameter (squeezing in the number fluctuations), a SMNc, condition from the Hillery{\&}Zubairy (TME) criterion~\cite{Hillery&ZubairyPRL2006}.

\section{Transformations among nonclassicality criteria: Examples} \label{sec:examples}
In this section, we present examples for obtaining a SMNc (quadrature-squeezing) condition  from a MPE (spin-squeezing) criterion~\cite{sorensen2001Nature}, in Sec.~\ref{sec:spin-squeezing}, and for obtaining a SMNc~(Mandel's $Q$ parameter) condition from a TME (Hillery{\&}Zubairy) criterion~\cite{Hillery&ZubairyPRL2006}, in Sec.~\ref{sec:QfromH&Z}. We also present derivation of Mandel's $Q$ parameter from Hillery{\&}Zubairy criterion, but this time, using beam splitter transformation in Sec.~\ref{sec:Q_withBS}.

\subsection{Single-mode squeezing from \\ spin-squeezing criterion} \label{sec:spin-squeezing}

S{\o}rensen {\it et al.}~\cite{sorensen2001Nature} introduced an inseparability criterion witnessing the entanglement of $N$ particles
\begin{equation}
\nu^2\equiv \frac{N(\Delta \hat{S}_{\bf n_1})^2}{\langle\hat{S}_{\bf n_2}\rangle^2 + \langle\hat{S}_{\bf n_3}\rangle^2} < 1.
\label{spinsqueezing}
\end{equation}
They demonstrate that for a general mixed separable many-particle state, given in Eq.~(\ref{SorensenEnt}),  $\nu$ is larger than 1. Hence, $\nu<1$ witnesses the many-particle entanglement, i.e. the many-particle state cannot be written as Eq.~(\ref{SorensenEnt}). In Eq.~(\ref{spinsqueezing}), $\hat{S_{\bf n}}$ are the operators for any orthogonal spin components, see Eq.~(\ref{HPtransform}).

One can choose $\hat{S}_{{\bf n}_1}=\hat{S}_x$ and $\hat{S}_{{\bf n}_2,{\bf n}_3}=\hat{S}_{y,z}$, where $\hat{S}_x=(\hat{S}_++\hat{S}_-)/2$, $\hat{S}_y=i(\hat{S}_+-\hat{S}_-)/2$. In the $N\to\infty$ limit, collective spin operators transform as  $\hat{S}_x\to\sqrt{N}(\hat{b}^\dagger+\hat{b})/2=\sqrt{N}\hat{x}/\sqrt{2}$ and $\hat{S}_y\to\sqrt{N}i(\hat{b}^\dagger-\hat{b})/2=\sqrt{N}\hat{p}/\sqrt{2}$. In this limit, $\hat{S}_z=\hat{b}^\dagger\hat{b}-N/2\to-N/2$. Thus, performing the cancellations, many-particle entanglement criterion~(\ref{spinsqueezing}) transforms to
\begin{equation}
(\Delta \hat{x}_b)^2 < 1/2,
\label{modesqz}
\end{equation}  
that is the well-known quadrature-squeezing condition for the single-mode field $\hat{b}$, with $\hat{x}_b=(\hat{b}^\dagger+\hat{b})/\sqrt{2}$.

Here, we show that a SMNc condition can be obtained from a MPE criterion. In Ref.~\cite{tasgin2017many}, we do the reverse. We anticipate one of the possible forms of a MPE criterion by examining the form of a Mandel's Q-parameter. The MPE criterion we obtain in Ref.~\cite{tasgin2017many} and the spin-squeezing criterion in Eq.~(\ref{spinsqueezing}) work good for states of different nature.  Hence, the method, we introduce in this subsection, can be used also for anticipating forms for MPE criteria.

%

\subsection{Mandel's Q-parameter from H{\&}Z criterion} \label{sec:QfromH&Z}

Next, we show that two-mode entanglement criteria can be practically transformed into single-mode nonclassicality criteria using many-particle representation as a bridge, as introduced in Sec.~\ref{sec:TMEtoMPE}. In particular, we show that Mandel's Q-parameter~\cite{mandel1979sub}, a SMNc criterion, can be obtained from the Hillery{\&}Zubairy~(a TME) criterion~\cite{Hillery&ZubairyPRL2006}.

A set of sufficient criteria for two-mode entanglement have been introduced by Hillery{\&}Zubairy~(H{\&}Z)~\cite{Hillery&ZubairyPRL2006} as
\begin{equation}
\left|\langle \hat{c}_g^m (\hat{c}_e^\dagger)^n \rangle\right|^2 > \left\langle (\hat{c}_g^\dagger)^m (\hat{c}_g)^m (\hat{c}_e^\dagger)^n (\hat{c}_e)^n \right\rangle,
\label{H&Znm}
\end{equation}
which witnesses the inseparability of a two-mode system for any integer values of $n$,$m$=1,2$\ldots$. $\hat{c}_g$ and $\hat{c}_e$ are annihilation operators for the two modes under consideration. One can obtain the Mandel's Q-parameter, i.e. $Q<1$ witnesses the single-mode nonclassicality, by setting $m=n=2$
\begin{equation}
\left|\langle \hat{c}_g^2 (\hat{c}_e^\dagger)^2 \rangle\right|^2 > \left\langle (\hat{c}_g^\dagger)^2 (\hat{c}_g)^2 (\hat{c}_e^\dagger)^2 (\hat{c}_e)^2 \right\rangle,
\label{H&Z2}
\end{equation}
as follows.

As demonstrated in Sec.~\ref{sec:TMEtoMPE}, for $N\to\infty$ a two-mode entangled state implies an inseparable many-particle system. Thus, a TME criterion implies a MPE criterion for $N\to\infty$. First, we express the two-mode operators in terms of collective spin operators of the many-particle system, e.g. $\hat{c}_e^\dagger \hat{c}_g=\hat{S}_+$. We note that such an expression can be obtained easily since the HZ TME criterion contains only particle number conserving terms, e.g. not terms like $\hat{c}_e\hat{c}_g$. Next, we perform a Holstein-Primakoff transformation on the MPE criterion which is valid merely for $N\to\infty$, e.g. $\hat{S}_-=\sqrt{N-\hat{b}^\dagger \hat{b}} \: \hat{b} \to \sqrt{N}\:\hat{b}$. Therefore, in sum, we obtain a SMNc condition from a TME criterion, using the many-particle system or representation as a bridge.

For the sake of obtaining the SMNc condition, we transform the term on the right hand side~(RHS) of Eq.~(\ref{H&Z2}) to
\begin{equation}
(\hat{c}_g^\dagger)^2 \hat{c}_g^2 (\hat{c}_e^\dagger)^2 \hat{c}_e^2 = 
\hat{c}_g^2 (\hat{c}_g^\dagger)^2  (\hat{c}_e^\dagger)^2 \hat{c}_e^2
-4\hat{c}_g \hat{c}_g^\dagger  (\hat{c}_e^\dagger)^2 \hat{c}_e^2
+2(\hat{c}_e^\dagger)^2 \hat{c}_e^2
\label{cgceidentitiy}
\end{equation} 
using the commutation relations $[\hat{c}_{e,g},\hat{c}_{e,g}^\dagger]=1$. The first term on the RHS of Eq.~(\ref{cgceidentitiy}) can be identified as $\hat{S}_+^2\hat{S}_-^2$. We work in the regime where $N$ is very large compared to  $\langle \hat{c}_e^\dagger \hat{c}_e\rangle$, i.e. the number of particles in the excited state. We notice that the first term on the RHS of Eq.~(\ref{cgceidentitiy}) is proportional to $N^2$ while the second and third terms are in the orders $N$ and $1$, respectively. Then, the last two terms can be neglected for $N\rightarrow\infty$. So, Eq.~(\ref{H&Z2}) takes the form
\begin{equation}
\left|\langle \hat{c}_g^2 (\hat{c}_e^\dagger)^2 \rangle \right|^2 > \langle  (\hat{c}_e^\dagger)^2  \hat{c}_g^2 \: \hat{c}_e^2 (\hat{c}_g^\dagger)^2   \rangle
=\langle \hat{S}_+^2\hat{S}_-^2 \rangle \; .
\end{equation}
The $\hat{S}_+^2\hat{S}_-^2$ term transforms to the desired form, $(\hat{b}^\dagger)^2\hat{b}^2$, after the Holstein-Primakoff transformation. The term on the left hand side~(LHS) can be related to the number operator $\langle\hat{b}^\dagger\hat{b}\rangle$, corresponds to the number of quasiparticle excitations~\cite{tasgin2017many}, using the Cauchy-Schwartz inequality
\begin{equation}
\langle \hat{c}_g^\dagger\hat{c}_e \: \hat{c}_e^\dagger\hat{c}_g \rangle
\langle \hat{c}_g\hat{c}_e^\dagger \: \hat{c}_g^\dagger\hat{c}_e \rangle \geq 
\langle \hat{c}_g^2 (\hat{c}_e^\dagger)^2 \rangle,
\end{equation}
where LHS can be written in terms of collective spin operators as $\langle \hat{S}_-\hat{S}_+\rangle\langle \hat{S}_-\hat{S}_+\rangle$ which becomes $\langle \hat{S}_+\hat{S}_-\rangle^2$ neglecting the term $\sim N$ compared to the $N^2$ term. Hence, Eq.~(\ref{H&Z2}) now becomes
\begin{equation}
| \langle\hat{S}_+\hat{S}_-\rangle|^2 >  \langle \hat{S}_+^2\hat{S}_-^2 \rangle\langle \hat{S}_+^2\hat{S}_-^2 \rangle
\label{Qspin}
\end{equation}
for $N$ sufficiently large. 

Thus, now, the TME criterion is expressed in terms of the many-particle, collective spin, operators for $N\to\infty$. Then, as the next step, we use the relation MPE$\to$SMNc, see Secs.~\ref{sec:MPEtoSMNc} and \ref{sec:HPtransform}, where an entangled many-particle state is shown to imply a nonclassical single-mode state as $N\to\infty$. Applying the Holstein-Primakoff transformation, summarized in Eq.~(\ref{HPtransform}), inequality~(\ref{Qspin}) becomes
\begin{equation}
| \langle \hat{b}^\dagger\hat{b} \rangle|^2 >  \langle (\hat{b}^\dagger)^2 \hat{b}^2 \rangle,
\label{MandelQb}
\end{equation}
that is the Mandel's Q-parameter~\cite{mandel1979sub} for a single-mode field. Mandel's Q-parameter can alternatively be expressed as
\begin{equation}
\langle(\Delta \hat{n}_b)^2\rangle < \langle\hat{n}_b\rangle,
\end{equation}
where number fluctuations reduce below the shot-noise limit. Here, $\hat{n}_b=\hat{b}^\dagger \hat{b}$.

Using the inequality set~(\ref{H&Znm}), one can also identify higher order nonclassicality conditions
\begin{equation}
| \langle \hat{b}^\dagger\hat{b} \rangle|^\ell >  \langle (\hat{b}^\dagger)^\ell \hat{b}^\ell \rangle \; ,
\label{nonclasssetHZ}
\end{equation}
other than the standard form Eq.~(\ref{MandelQb}), for higher-order correlation functions $g^{(\ell)}$~\cite{ScullyZubairyBook}, with $\ell$ integer. In addition to set of inequality (\ref{nonclasssetHZ}), one also obtains
\begin{equation}
|\langle \hat{b}\rangle|^2 > \langle \hat{b}^\dagger\hat{b}\rangle
\end{equation}
for $m=n=1$ in Eq.~(\ref{H&Znm}), which is forbidden by the Cauchy-Schwartz inequality to satisfy.

Unfortunately, two entanglement criteria, Simon-Peres-Horodecki~(SPH)~\cite{SimonPRL2000} and Duan-Giedke-Cirac-Zoller~(DGCZ)~\cite{DGCZ_PRL2000}, ---which are both necessary and sufficient criteria for Gaussian states--- require the evaluation of non-number conserving terms  like $\langle\hat{c}_g\hat{c}_e\rangle$. Hence, this kind of criteria cannot be transformed into single-mode nonclassicality conditions using the Holstein-Primakoff approach. For this reason, in Ref.~\cite{tasgin2015measure} we use the beam-splitter approach to deduce the degree of single-mode nonclassicality from Simon-Peres-Horodecki criterion~\cite{SimonPRL2000,adesso2004extremal,Vidal:02}. One may also expect a further connection between the entanglement depth~\cite{Sorensen&MolmerPRL2001entanglement} in many-particle inseparability and the degree of single-mode nonclassicality (of quasiparticles) due to Holstein-Primakoff transformation.

\subsection{Mandel's Q-parameter using BS approach} \label{sec:Q_withBS}

Additionally, we show that Mandel's Q-parameter condition can alternatively be obtained using the beam splitter formalism~\cite{hillery2006PRA,tasgin2015measure} using a ``lower-order" Hillery{\&}Zubairy~(H{\&}Z) criterion. One could naturally raise the question: if the two methods. i.e. TME$\to$MPE$\to$SMNc and TME$\xrightarrow{\rm BS}SMNc$, have somehow a link in between implicitly, that is not apparent to the naked eye at first glance? Our observation, that the two methods achieve the Mandel's Q-parameter using HZ criteria of not the same order, however, reduces the possibility for the existence of such a link.

One can also determine the nonclassicality of a single-mode field according to the two-mode entanglement it creates at the output of a beam splitter. The input beam is mixed either with vacuum noise or a coherent state in order to guarantee that the entanglement generated at the output of the beam splitter originates from the nonclassicality of the input beam~\cite{Kim:02,Asboth:05,Vogel:14}. Input modes are transformed~\cite{Kim:02,Kim:02,Campos:89,aharonov1966quantum} via the beam splitter operator
\begin{equation}
\hat{B}(\beta)=\hat{D}_{\rm \scriptscriptstyle TM}=e^{ \beta\hat{a}_2^\dagger \hat{a}_1 - \beta^*\hat{a}_1^\dagger \hat{a}_2 } \; .
\label{BSoperator}
\end{equation}
One acts  $\hat{B}(\beta)$ on the initially separable state of two-mode state,
\begin{equation}
|\psi_{12}\rangle = |\psi_a\rangle_1 \otimes |0\rangle_2,
\end{equation}
where $|\psi_a\rangle$ is the state of the single-mode field ($\hat{a}$) incident to the beam splitter, whose nonclassicality is to be determined, and $|0\rangle_2$ is the vacuum field mixing with $\hat{a}$. In the language of operators, $\hat{a}$ is the single-mode field operator which is the initial to the $\hat{a}_1$ operator to be transformed in Eq.~(\ref{BSoperator}). The vacuum input is the initial of the $\hat{a}_2$ operator.

In the language of states, Schrodinger picture, $|\psi_a\rangle=\sum_{n=0}^\infty c_n |n\rangle $ is the initial state of the first mode which transforms  via Eq.~(\ref{BSoperator}). The initial state of the second mode, on which $\hat{a}_2$ operator acts, is vacuum. This method is equivalent to $|\psi_{12}\rangle=f(\mu_1\hat{a}_1^\dagger+\mu_2\hat{a}_2^\dagger)|0\rangle_1\otimes|0\rangle_2$ transform~\cite{aharonov1966quantum} on the two-mode wave function, where function $f$ is defined by the single-mode wave function $|\psi_a\rangle=\left( \sum_{n=0}^\infty d_n (\hat{a}^\dagger)^n\right)|0\rangle_a$ as $f(z)=\sum_{n=0}^\infty d_n z^n$. Here, $\mu_1=te^{i\phi}$ and $\mu_2=r$  are the coefficients in the well-known form~\cite{Kim:02,Campos:89},
\begin{subequations}
\begin{eqnarray}
\hat{a}_1(\beta)=\hat{B}^\dagger(\beta)\hat{a}_1\hat{B}(\beta)=te^{i\phi}\hat{a}_1(0) + r\hat{a}_2(0) \; ,
\label{BStransa}
\\
\hat{a}_2(\beta)=\hat{B}^\dagger(\beta)\hat{a}_2\hat{B}(\beta)=-r\hat{a}_1(0) + te^{-i\phi}\hat{a}_2(0) \; ,
\label{BStransb}
\end{eqnarray}
\end{subequations}
of the beam splitter transformation, in the language of operators, i.e. the Heisenberg picture. $t^2$ and $r^2$ correspond to transmission and reflection coefficients, and $\phi$ is the phase of the beam splitter. We note that $\hat{a}_1(0)=\hat{a}$ where $\hat{a}$ refers to the input operator to be examined. $\hat{a}_2(0)$ refers to the second input (vacuum) mode. As usual, it is easier to work in the Heisenberg picture, where operators transform.

An expectation value including $\hat{a}_1$, $\hat{a}_2$ operators, for instance,
\begin{equation}
\langle \hat{a}_1\hat{a}_2^\dagger\rangle = {}_2\langle0| \otimes {}_1 \langle\psi_a| \hat{B}^\dagger(\xi) \hat{a}_1 \hat{a}_2^\dagger \hat{B}(\xi) |\psi_a\rangle_1 \otimes |0\rangle_2 \; ,
\end{equation}
can be evaluated to
\begin{eqnarray}
\langle \hat{a}_1\hat{a}_2^\dagger\rangle = {}_2\langle0| {}_1 \langle && \psi_a| (te^{i\phi}\hat{a} + r\hat{a}_2(0)) 
\\ \nonumber
&& \times (-r\hat{a}^\dagger + te^{i\phi}\hat{a}_2^\dagger(0)) |\psi_a\rangle_1 |0\rangle_2 \; ,
\end{eqnarray}
performing the transformations (\ref{BStransa}) and (\ref{BStransb}), where $\hat{a}_2(0)$ is the not-evolved input operator referring to the vacuum noise.

In order to determine if the input state $|\psi_a\rangle_1$ possesses SMNc, we check if the beam splitter generates entanglement at the output, by calculating the HZ criterion
\begin{equation}
|\langle \hat{a}_1 \hat{a}_2^\dagger\rangle|^2 > \langle \hat{a}_1^\dagger \hat{a}_1 \hat{a}_2^\dagger \hat{a}_2  \rangle \; .
\label{H&Z1}
\end{equation}
Using the transformations (\ref{BStransa}) and (\ref{BStransb}) one can relate the terms in Eq.~(\ref{H&Z1}) to the ones for single-mode field ($\hat{a}$) as
\begin{eqnarray}
\langle \hat{a}_1 \hat{a}_2^\dagger\rangle =-rte^{i\phi} \langle \hat{a}^\dagger \hat{a} \rangle,
\label{a1a2BS1}
\\
\langle \hat{a}_1^\dagger \hat{a}_1 \hat{a}_2^\dagger \hat{a}_2  \rangle = t^2r^2 \langle \hat{a}^\dagger \hat{a}^\dagger \hat{a} \hat{a}  \rangle,
\label{a1a2BS2}
\end{eqnarray}
which simply give the condition for the single-mode nonclasivality of the input mode $\hat{a}$ as
\begin{equation}
|\langle\hat{a}^\dagger\hat{a}\rangle|^2 > \langle \hat{a}^\dagger \hat{a}^\dagger \hat{a} \hat{a}\rangle \; .
\label{MandelQBS}
\end{equation}

Unlike the TME$\to$MPE$\to$SMNc approach, introduced in Secs.~\ref{sec:TMEtoMPE} and \ref{sec:QfromH&Z}; using the beam splitter approach, it is possible to obtain single-mode nonclassicality conditions from TME criteria involving particle number non-conserving terms~\cite{hillery2006PRA,tasgin2015measure}. However, we could not manage to obtain an easy analytical form for a SMNc condition from the TME criterion introduced by Duan-Giedke-Cirac-Zoller~(DGCZ)~\cite{DGCZ_PRL2000}. In the case of H{\&}Z criterion, the coefficients $r$, $t$ and $e^{i\phi}$ could be cancelled in Sec.~\ref{sec:QfromH&Z}. However, in the DGCZ case numerical minimization with respect to $r$ and $\phi$ is required~\cite{Asboth:05}. Still, we anticipate the SMNc condition, could be obtained from DGCZ TME criterion, as the quadrature-squeezing condition.

\section{Summary and Discussions} \label{sec:conclusions}

In summary, we study different representations of a symmetric many-particle system of two-level atoms in the large particle number limit $N\to\infty$. We utilize these connections for obtaining nonclassicality criteria from one another, i.e. among many-particle, two-mode and single-mode systems. In particular, we obtain SMNc conditions both from many-particle and two-mode entanglement criteria. 

An ensemble of $N$ identical and indistinguishable atoms can occupy the (exchange) symmetric Dicke states~(see figure 1 in Ref.~\cite{dicke1954SR}). Symmetric Dicke states can be represented both by single-mode and two-mode operators/states. In particular, Dicke (number) states and atomic coherent states converge to single-mode Fock number states and coherent states, respectively, at the large particle limit. This map can be used to connect the entanglement of a symmetric many-particle state to the nonclassicality of a single-mode field. Hence, an entangled many-particle state implies a nonclassical state in this mapping~\cite{tasgin2017many}. Therefore a many-particle entanglement criterion can be used as a single-mode nonclassicality condition in this limit. Since the states are experimentally not accessible in most cases, one maps/transforms the operators alternatively.  A many-particle entanglement criterion can be transformed into single-mode nonclassicality condition via Holstein-Promakoff transformation, e.g., $\hat{S}_-=\sqrt{N-\hat{b}^\dagger \hat{b}}\:\hat{b}$ which becomes $\hat{S}_-\to \sqrt{N}\hat{b}$ in the infinite particle limit. $\hat{b}$ accounts the quasiparticle excitations of the many-particle system. We demonstrate that spin-squeezing (many-particle entanglement) criterion becomes the quadrature-squeezing condition in this limit.

One can also represent a symmetric many-particle system in terms of two-modes, e.g. $\hat{S}_+=\hat{c}_e^\dagger\hat{c}_g$ where $\hat{c}_{g,e}^\dagger$ creates an identical particle in the ground/excited state. Here, $\hat{c}_e$ actually is equivalent to $\hat{b}$ operator of the single-mode representation. We notice that separable (many-particle) atomic coherent states~(ACSs) are generated via $e^{\xi\hat{S}_+-\xi^*\hat{S}_-} \equiv e^{\xi\hat{c}_e^\dagger\hat{c}_g-\xi^*\hat{c}_g^\dagger\hat{c}_e}\equiv e^{\alpha\hat{b}^\dagger-\alpha^*\hat{b}}$ and beam splitter operator conserves the nonclassicality in the two-mode representation. (Here, $\alpha=\sqrt{N}\xi$.) This makes us introduce a connection also between two-mode entanglement~(TME) and and many-particle entanglement~(MPE). 

%

We use TME$\to$MPE connection to develop a a method for deriving SMNc conditions from TME criteria, i.e. following the path TME$\to$MPE$\to$SMNc. This method is an alternative to beam splitter approach, which is also used for TME$\to$SMNc derivation, and demonstrates the stunning beauty of the transformations among three different systems, see Sec.~\ref{sec:QfromH&Z}. The new method, however, is limited with the number conserving terms only. Two well-known entanglement criteria~\cite{SimonPRL2000,DGCZ_PRL2000}, for instance, cannot be used to obtain a SMNc condition with the new method, since they include particle nonconserving terms. This connection can also be used to anticipate a form for a nonexisting many-particle entanglement criteria.

We believe that the intriguing connections, we demonstrate in this paper, will stimulate new research on the intersection of many-particle, single-mode and two-mode systems.

\begin{acknowledgements}
We thank \"{O}zg\"{u}r E. M\"{u}stecapl{\i}o\u{g}lu and G\"{u}rsoy B. Akg\"{u}\c{c} for illuminating and leading discussions. We acknowledge  support  from TUBITAK-3501  Grant No.  112T927,  TUBITAK-1001  Grant Nos.  114F170 and 117F1118, TUBA-GEBIP 2017 and Hacettepe University BAP Grant No: FBI-2018-17423.
\end{acknowledgements}

\bibliography{D:/AAADocuments/latexbib/bibliography}

\end{document}